\documentclass{bioinfo}

\copyrightyear{2015} \pubyear{2015}

\access{Advance Access Publication Date: Day Month Year}
\appnotes{Manuscript Category}

\usepackage[]{algorithm2e}
\usepackage{amsmath,amssymb}

\usepackage{color}

\begin{document}
\firstpage{1}

\subtitle{Genome Analysis}

\title[BOAssembler]{BOAssembler: a Bayesian Optimization Framework to Improve RNA-Seq Assembly Performance}
\author[Mao and Jiang \textit{et~al}.]{Shunfu Mao\,$^{\text{\sfb 1}*}$, Yihan Jiang\,$^{\text{\sfb 1}*}$, Edwin Basil Mathew\,$^{\text{\sfb 2}}$ and Sreeram Kannan $^{\text{\sfb 1}}$}

\address{
$^{\text{\sf 1}}$Department of Electrical and Computer Engineering, University of Washington, Seattle, USA\\
$^{\text{\sf 2}}$Data Science Program, University of Washington, Seattle, USA}

\corresp{$^\ast$ Equal contribution, to whom correspondence should be addressed.}

\history{Received on XXXXX; revised on XXXXX; accepted on XXXXX}

\editor{Associate Editor: XXXXXXX}

\abstract{\textbf{Motivation:} 
%DNA/RNA assemblers serve as the powerhouse for Bioinformatics revolution. As DNA for now cannot be read in a single long strand with high enough accuracy, the performance of assembler is critical and inevitable for all downstream applications. Hyper-parameters of assembler are critical and decisive to enable further applications. The assemblers typically suggests some default hyper-parameter, which doesn't guarantee a good performance on other species and environment. Hand-tuned hyper-parameters, based on heuristic and luck, doesn't show consistent improvement when facing practical dataset.
High throughput sequencing of RNA (RNA-Seq) can provide us with millions of short fragments of RNA transcripts from a sample. How to better recover the original RNA transcripts from those fragments (RNA-Seq assembly) is still a difficult task. For example, RNA-Seq assembly tools typically require hyper-parameter tuning to achieve good performance for particular datasets. This kind of tuning is usually unintuitive and time-consuming. Consequently, users often resort to default parameters, which do not guarantee consistent good performance for various datasets.
\\
\textbf{Results:}
%In this paper, we propose a data-driven machine-learning method: Bayesian Optimization (BO) to optimize assembler in a black-box fashion. With simple and effect data-driven approach, the precision and recall of assembler increases dramatically comparing to using default settings as is. BO method boosts the performance on assembly level, thus the improvement can be further propagated to later downstream applications, which shows the potential to be the standard for future Bioinformatics pipeline.\\
Here we propose BOAssembler, a framework that enables end-to-end automatic tuning of RNA-Seq assemblers, based on Bayesian Optimization principles. Experiments show this data-driven approach is effective to improve the overall assembly performance. The approach would be helpful for downstream (e.g. gene, protein, cell) analysis, and more broadly, for future bioinformatics benchmark studies.
\\
\textbf{Availability:} \href{https://github.com/olivomao/boassembler}{https://github.com/olivomao/boassembler}
\\
\textbf{Contact:} \href{shunfu@uw.edu}{shunfu@uw.edu}, \href{yihanrogerjiang@gmail.com }{yihanrogerjiang@gmail.com}\\
\textbf{Supplementary information:} Supplementary data are available at \textit{Bioinformatics}
online.}

\maketitle

\section{Introduction}

%reference assembly
Sequence assembly is a process to recover the original genomic sequences from their sampled reads. Based on sequencing technology (DNA/RNA) and model genome availability, there are different assembly problems. In this study, we focus on reference-based RNA-Seq assembly, which is the first step to understand gene, protein and cell functions.

Existing reference-based RNA-Seq assemblers (such as Cufflinks \cite{Cufflinks}, Stringtie \cite{Stringtie}) usually align reads onto reference genome first, and based on the alignment build a graph where each node represents a genome region (exon) and each edge represent two node regions are aligned by some reads. They then traverse the graph to find paths as recovered RNA transcripts. 

The assembly problem is essentially NP-hard \cite{kececioglu_combinatorial_1995} and existing tools resort to heuristic methods. For example, from the graph, Stringtie will extract heaviest paths iteratively. These methods usually require parameter tuning to achieve good performance for particular datasets. Since most users may not understand the meaning of the parameters well and tuning is tedious and time-consuming, they usually end up with default settings. An automatic tuning framework, therefore, is necessary.

%baysian optimization
In machine learning (ML), Bayesian Optimization (BO) is gaining a surge of interest as its usefulness in tuning hyper-parameters for modern deep learning systems \cite{Snoek:2012}. BO is favorable for optimizing objective functions that are expensive to evaluate and are over continuous domains of less than 20 dimensions \cite{Frazier2018}. BO has become widely used in most deep learning systems such as Natural Language Processing (NLP) \cite{bo_nlp}, Reinforcement Learning (RL) \cite{bo_gaming}, and Channel Coding \cite{bo_cc}. Depending on algorithms and programming languages, several popular BO packages have been developed, such as GPyOpt \cite{gpyopt2016}.

%related works
There are limited work to introduce BO into computational biology fields. Recently \cite{Quitadadmo2017} applies BO to improve eQTL analysis. To the best of our knowledge, no work has introduced BO to assembly tasks yet, which are fundamentally graph problems with their own unique challenges. 
To fill this gap, we have developed BOAssembler, which is a framework able to incorporate existing assemblers (such as Stringtie) and BO methods (such as GPyOpt) to assist assembler developers and biologists to spend minimal efforts to have the assemblers' hyper-parameters automatically fine tuned for particular datasets.

%contribution
Our contribution as follows:
\begin{itemize}
\item We firstly explore the BO methods in (reference-based RNA-Seq) assembly tasks.
\item Our designed experiments show that BO is overall effective to improve assembly.
\item An open source end-to-end framework (BOAssembler) is provided for the assembly community to use.

\end{itemize}

% Yihan: I prefer to talk about assembler first
%\begin{methods}
\section{Methods}
In this section we first introduce assemblers, and then explain Bayesian Optimization, and finally describe our BOAssembler framework that combines both.

\subsection{Assembler}
%assembly
There are two kinds of RNA-Seq assembly problems: de novo assembly and reference-based assembly. For de novo assembly, we only have RNA-Seq reads, which is common in non-model organisms. For reference-based assembly, there is additional knowledge on the genome of the organism. De novo assembly is appearantly more challenging and typical tools (such as Trinity \cite{Grabherr2011} and recently Shannon \cite{Kannan2016}) require much more computational resources and more complicated evaluations. As a first step to bridge assembly and BO, we focus on reference-based RNA-Seq assembler. In particular, we focus on the widely used Stringtie%\footnote{We've also included another popular assembler Cufflinks into BOAssembler, but it takes much longer time even for smaller dataset.}
, as recommended in \cite{Hayer2015}.

A typical reference-based RNA-Seq assembly includes aligning sampled RNA-Seq reads onto a reference genome using external tools such as STAR \cite{Dobin2012} etc. For Stringtie, a (splice) graph will be prepared where each node represents a unique exonic region supported by aligned reads and edges indicate how nodes are bridged by reads. Graph traversal algorithms will be applied to find paths as transcripts to best explain the constraints from graph nodes and edges.

%issues of tuning parameters 
Since assembly problems are NP hard \cite{kececioglu_combinatorial_1995}, existing algorithms take a lot of heuristics (bunch of thresholds), assembler performance depends heavily on hyper-parameters. 
%\textbf{add some example of parameters here to give audiance a hint of what parameter is:}
For example in Stringtie, parameter '-f' sets a fractional threshold below which predicted transcripts will be discarded, and a lower value encourages transcripts to be retained to improve sensitivity.

Developers of assemblers typically tune parameters by intuition on a few datasets, and offer default parameters for assembler users to use. As assembler's performance for various datasets are usually parameter dependent, a more systematic method of tuning parameter is needed. Parameters are continous, and not of low dimensions, which makes grid search on all possible combination prohibitive. Random search \cite{bergstra:2012} are expensive to guarantee good coverage, which is not favorable to tune parameter for assembler. We propose BO based method, which is a systematic parameter tuning method with limited number of evaluations in the following part.

\subsection{Bayesian Optimization}\label{sec:bo}
% abstraction of assembler function
The reference-based assembler together with its evaluation can be represented as an abstract function $f(D, \theta)$, where $D$ includes both read alignment for assembly and reference transcriptome (a set of ground truth RNA transcripts) for evaluation, and $\theta$ is the parameters of the abstract function with parameters of dimension $d$. After read alignments are assembled with given parameter $\theta$, the assembly output (a set of RNA transcripts) will be compared with the reference transcriptome, and the quality of assembly is measured with scalar metrics such as precision $p$ and sensitivity $s$. $f(D, \theta)$ outputs a score based on $p$ and $s$. Our goal is to find a global optimal $\theta$ which maximizes $f(D, \theta)$, with limited evaluations of $f(D, \theta)$ since running assembler is time consuming. 

% Intro of BO
BO aims at maximizing a real-value black-box function $f(\theta)$ with respect to $\theta$~\cite{frazier:2018} in a gradient-free approach (here $D$ is fixed). BO consists of a statistical surrogate objective function to model the input-output relationship between $\theta$ and $f(\theta)$, and an acquisiton function to decide what to sample next. Firstly BO evaluates randomly chosen $K$ datapoints of $\theta$, and fits the prior statistical objective model. Then BO iteratively updates the posterior model with newly acquired $f(\theta_k)$, and selects $\theta_{k+1}$ to evaluate according to posterior. BO is a systematic approach to explore the parameter space according to a Bayesian model with limited allowed evaluations.

Gaussian Process (GP) with Matern Kernel \cite{Snoek:2012} is a natural model for statistical objective function. Expected Improvement (EI) is a commonly used acquisition function. Our BOAssembler uses GP with Matern Kernel and EI as our primary BO method. For details, please refer to Supplementary Section 2. The procedure of iterative update, based on GP and EI, is described in Algorithm \ref{algo:bo_ei}.

%The BO algorithm for assembler $f(D, \theta)$ is described in Algorithm \ref{algo:bo_ei}, with $K$ datapoints for BO initialization, and $T$ iterations to acquire.

\vspace*{-10pt}
\begin{algorithm}
 \KwData{$D$, $K$, $T$}
 \KwResult{Best parameter $\theta^{*}$}
 Fit the GP with $K$ initial samples $\theta_k, k\in \{1,...,K\}$;\\
 i=0;\\
 \While{i<T}{
  Update the GP posterior probability distribution on $f$ using all availabel data;\\
  Use EI to compute the $\theta'$ with updated posterior distribution;\\
  Obtain $f(\theta')$;\\
  i++;\\
 }
 Return $\theta^{*}$ with best performance;\\
 \caption{Baysian Optimization with Expected Improvement}
 \label{algo:bo_ei}
\end{algorithm}
\vspace*{-15pt}

\subsection{Combine BO and Assembler}

%\subsubsection{BO is favorable for assembler parameter tuning}
\subsubsection{The Motivation to Combine}

Bayesian optimization works well for black-box gradient-free global optimization with moderate dimensionality. It is favorable to apply BO to optimize assembler parameters due to the following reasons:

\begin{itemize}
    \item Empirically BO works well for parameters with moderate dimensionality (less than 20). This is consistent with assemblers which typically have moderate number of parameters. The usage of BO inherently assume the parameters form a Gaussian Process.
    
    \item $f(.)$ is continuous, and the parameter $\theta$ are correlated, and has well-defined feasible set. This meets the requirements of BO.

    %a bit ambiguous for me
    %{\cn This is typically required to model $f(.)$ using GP.}
    
    \item The function $f(.)$ %$f(D, \theta)$
    is expensive to evaluate, thus to evaluate all possible combinations of parameters is prohibitive.
    
    \item $f(.)$ is a 'black-box', while gradient-based optimization methods cannot be applied. Assemblers typically do not have a gradient due to usage of thresholds, which makes black-box method favorable.
    
\end{itemize}

A major drawback of BO is its inference time grows cubically with respect to number of iterations~\cite{snoek:2015}. For assembler the scalability issue is not severe, since we observe convergence typically at $40$ to $50$ iterations, and the major bottleneck is the assembly part. %evaluating the assembler: comparing to the computational resource of assembler, the resource spent on BO is ignorable.

\subsubsection{The Architecture}

\begin{figure}[!h] %[!tpb]%figure1
%\fboxsep=0pt\colorbox{gray}{\begin{minipage}[t]{235pt} \vbox to 100pt{\vfill\hbox to
%235pt{\hfill\fontsize{24pt}{24pt}\selectfont FPO\hfill}\vfill}
%\end{minipage}}
\centerline{\includegraphics[width=0.5\textwidth]{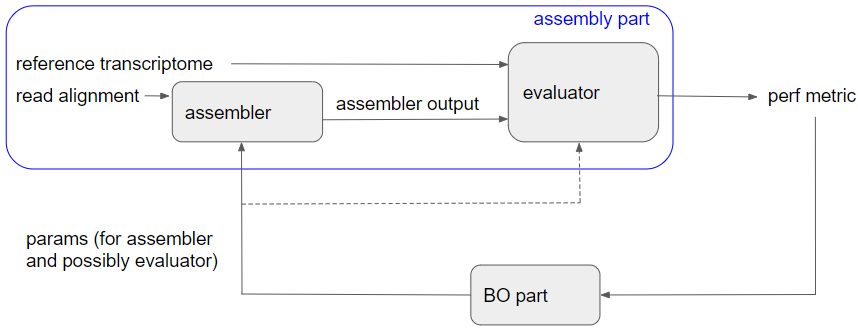}}
\caption{BOAssembler architecture.}\label{fig:architecture}
\end{figure}

Figure \ref{fig:architecture} illustrates the overall architecture of BOAssembler. There are two parts: the assembly part (e.g. $f(D, \theta)$) and BO part.

The assembly part wraps up the RNA-Seq reference-based assembler (here Stringtie), which takes fixed read alignment as well as adjustable assembler parameters as input, and outputs assembled RNA transcripts (in gtf format). In addition, the assembly part includes an evaluator block to access the assembly output. Basically, it calls the gffcompare\footnote{https://ccb.jhu.edu/software/stringtie/gffcompare.shtml} tool, which takes as input the assembly output and reference transcriptome, and outputs sensitivity and precision statistics. The sensitivity is the percentage of reference RNA transcripts that have been correctly recovered, and the precision means the percentage of assembled transcripts that correctly match the reference transcriptome. We further combine the sensitivity and precision (such as F1 score) as $f(D,\theta)$, to be used by the BO part. The evaluator may also take adjustable parameters as discussed in Section \ref{sec:metric}.

The BO part has its theory described in Section \ref{sec:bo}. More concretely, it wraps up existing BO methods (such as GPyOpt). It treats the assembly part as a black box, where the input to the box is the parameters for assembler and evaluator, and the output of the box is the combined performance metric for the assembler (such as F1 score). The BO part will iteratively optimize the parameters for the black box function (e.g. the assembly part) based on the feedback of performance metric.

\subsubsection{Metrics to optimize}\label{sec:metric}

The assembly part outputs $f(D,\theta)$, which is a metric score and serves as an input to BO part. In particular, it is defined as a weighted F1 score ($S_w=\frac{\lambda p \times (1-\lambda)s}{\lambda p + (1-\lambda)s}$) on top of the evaluator's output in terms of sensitivity $s$ and precision $p$ of the assembly. $\lambda \in (0,0.5)$ is also BO tunable.

There are several candidate metrics including the mean value ($S_m=\frac{s+p}{2}$) and the F1 score ($S_{F1}=\frac{2 s p}{s+p}$). We found the BO part tends to overfit either $s$ or $p$ towards $1$ when using $S_m$. Though $S_{F1}$ is able to balance $s$ and $p$, we find $S_w$ is better to improve the final performance of sensitivity and precision. In our experiments, $s$ tends to have a lower value range than $p$ (due to many reference RNA transcripts do not have enough coverage), we hope to reward more for sensitivity improvement but still have gain on precision. Therefore, we come up with the weighted F1 score $S_w$, which uses BO to figure out how much percentage we want to reward especially for the improvement of sensitivity.   

%Finally the hyperparameter
%Discuss about the effect of hyper-parameters, for different user cases.
\subsubsection{Hyper-parameters}\label{sec:param}

Hyper-parameters are applied to assembler and to evaluator ($\lambda$). We focus on numerical (int, float) types. Each hyper-parameter has its name (e.g. '-f'), type (e.g. float), default value (e.g. $0.1$) and range (e.g. $(0.0,1.0)$). See Supplementary Section 3 for Stringtie's example. 
 
\subsubsection{Usage and Extension of BOAssembler}

To tune parameters for an existing assembler in BOAssembler, the user only needs to provide a small sample of read alignment which can be done by our provided scripts (see Supplementary Section 1). After some iterations, BOAssembler will report suggested parameters and its tuning history (e.g. per iteration's parameter and metrics) in a log file.

BOAssembler currently uses Stringtie as its default assembler. It supports Cufflinks as well. Extension to use other reference-based RNA-Seq assemblers is also straight forward. The user only needs to follow the Stringtie example, to add a line of Python code in a specified Python file, and to add config file according to our pre-defined format as mentioned in Section \ref{sec:param} to include the parameters to be tuned. 

\section{Result and Discussion}

\subsection{Datasets}
%data preparation
Our goal is to use BOAssembler to tune assembler's hyper-parameters on a smaller dataset, and apply recommended hyper-parameters on a large assembly task. Since the smaller dataset has representitive data of large assembly task, we expect tuned hyper-parameters can overall improve the large assembly task in terms of sensitivity and precision.

We build our results based on simulated datasets, since real datasets lack ground truth and it is hard to judge if an assembled RNA transcript is a false positive, or a new RNA transcript that has yet to be discovered. The simulated datasets are generated based on real ones.

Firstly we prepare three real datasets, including: 132.05M Illumina single end reads (50-bp) sampled from human embryonic stem cells (HESC) (GSE51861, used in \cite{ww_data}), 115.36M Illuminar pair end reads (101-bp) sampled from Lymphoblastoid cells (LC) (SRP036136, used in \cite{snyder_data}), and 183.53M Illuminar pair end reads (100-bp) sampled from HEK293T (Kidney) cells (SRX541227), previously produced and studied in StringTie \cite{Stringtie}.

Secondly, we use RSEM \cite{Li2011} to generate simulated reads from real datasets. To begin with, we choose LC reference transcripts (containing $207266$ RNA transcripts) as the ground truth reference transcriptome annotations. We then do quantification of real datasets using RSEM and get learned statistics from real datasets. Based on learned statistics, we use RSEM to sample simulated reads from ground truth reference transcriptome. The simulated HESC has 150M 50-bp single-end reads, simulated LC has 150M 101-bp pair end reads and simulated Kidney has 150M 100-bp pair end reads. 

Lastly, we use STAR \cite{Dobin2012} (2-pass strategy) to align three simulated datasets onto human reference genome (hg19)\footnote{http://hgdownload.cse.ucsc.edu/goldenpath/hg19/bigZips/}. From each alignement (in bam format), we subsample to get smaller alignment files of chromosome15 as fixed datasets for BOAssembler. The small datasets are about $1.5\%$, $3.1\%$, and $2.1\%$ of large datasets for HESC, LC and Kidney respectively. We've proposed another more complicated sampling method across chromosomes, which offer similar performance as discussed in Supplementary Section 1.

\subsection{Procedure}

For each dataset, we run BOAssembler on the smaller datasets. The evaluation for metric also uses a subset (e.g. chromosome 15) of reference transcriptome. Each iteration takes around 1 minute, and we typically see convergence of metric score around 40 to 50 iterations. Compared to grid search for possible combinations of 10 to 20 parameters, BOAssembler is much more efficient.

After automatic tuning, BOAssembler will recommend parameters with high metric scores. We then apply these parameters on large datasets, which typically take several hours to finish the assembly tasks using 25 cores of a linux server.

\subsection{Experiment Results}

\begin{table}[!h]
\processtable{Performance on Different Simulated Datasets\label{Tab:res}} {
\begin{tabular}
{@{}lllllll@{}}\toprule
{} & Sens & {} & Prec & {} & F1 & {}\\
{Dataset} & Def & Tune & Def & Tune & Def & Tune\\\midrule
HESC (small) & 22.1	&	39	&	31.9	&	59.3	&	26.11	&	47.05
\\
HESC (large) & 14.3	&	20.8	&	54.2	&	86.4	&	22.63	&	33.53
\\\midrule
LC (small) & 25.8	&	27	&	40	&	53.1	&	31.37	&	35.8
\\
LC (large) & 15.7	&	14.5	&	64.3	&	74.3	&	25.24	&	24.26
\\\midrule
Kidney (small) & 20.1	&	23.3	&	27.9	&	33.4	&	23.37	&	27.45
\\
Kidney (large) & 14.8	&	15.1	&	54.1	&	54.7	&	23.24	&	23.67
\\\botrule
\end{tabular}
}
{}%footnote
\end{table}
%\vspace{-0.1in}

%res summary
Table \ref{Tab:res} compares the performance of default parameters (Def) and BOAssembler-tuned parameters (Tune) for each simulated dataset, in terms of sensitivity (Sens), precision (Prec). We also list their standard F1 score here since it's related to the metric BOAssembler tries to optimize. But we'll focus on sensitivity and precision which are of practical interest.

As Table \ref{Tab:res} shows, BOAssembler has improved sensitivity, and precision for all small datasets. In particular, HESC small is improved by $16.9\%$ in sensitivity and $27.4\%$ in precision, LC samll is improved by $1.2\%$ in sensitivity and $13.1\%$ in precision, Kidney small is improved by $3.2\%$ in sensitivity and $5.5\%$ in precision. Notice that the real Kidney dataset has been used in Stringtie's original work, so the default parameters of Stringtie should have been adjusted for this dataset statistics. Still BOAssembler improves the its performance further.

The trend of performance improvement is mostly reflected in assembly tasks on large datasets, which is most interesting to us. In particular, HESC large is improved by $6.5\%$ in sensitivity and $32.2\%$ in precision, Kidney large is improved by $0.3\%$ in sensitivity and $0.6\%$ in precision. LC large has a small loss around $1.2\%$ in sensitivity, but it gains $10\%$, which is significant, in precision. The experiments show that by tuning hyper-parameters through BOAssembler on small datasets, we are able to improve large assembly tasks overall (though there could be fluctuations) to a smaller extent.

The diminished performance gain of tuned parameters on large datasets, compared to the gain on small ones, may be because of an averaging effects across more variant alignment statistics in large datasets. To better catch up large dataset statistics, we have also prepared small datasets selected from certain regions, the performance improvement trend is similar (see Supplementary Section 1).

%res interpretation
By comparing the BOAssembler suggested parameters with assembler's default ones, we could also gain more insights into the datasets. For example, in HESC small datasets, the parameter 'f' is suggested to decrease from $0.1$ to $0$, this will allow more transcripts of low expression levels to also be considered as assembly output (hereby improve sensitivity). Meanwhile, the parameter 'm' is suggested to increase from $200$ to $500$ to allow only longer (e.g. at least $500$) assembled transcripts to be considered (hereby improve precision).

\subsection{Discussion}
%discussion - benefits
We expect our study and developed BOAssembler will contribute to the assembly community as follows:
\begin{itemize}
\item For bioinformaticians who develop assembly algorithms, the framework or ideas behind it could provide them with more convenient ways to set default parameters for their assemblers.
\item For biologists who use reference-based RNA-Seq assemblers, BOAssembler can help them improve assembly performance, so they can gain better insights into the datasets, and the improved assembled RNA transcripts will be helpful for downstream gene, protein and cell related analysis.
\item For benchmark work of assemblers, %instead of hosting a challenge and inviting different teams to optimize their methods for the target datasets,
typically several datasets are prepared and different assemblers are compared for their default parameters. BOAssembler or its ideas will help the benchmark work in a fairer basis, since default parameters can not ganrantee consistent good performance across various datasets.
\end{itemize}

%discussion - limits
Whereas this is, to our best knowledge, the first efforts to bring assembly and BO together, there are interesting directions future directions.

\begin{itemize}
\item As from experiments, we have observed that the gain of tuned parameters gets diminished for larger datasets, which implies BO tuned parameter overfits to small training dataset. The idea of using additional validation dataset is shown in Supplementary Section 2.4. Since evaluating assembler is expensive, more efficient data subsampling and cross-validation methods to avoid overfitting are an interesting future direction.

%This could be potentially improved if we pick smaller datasets from different chromosomes, and optimize them in a joint way.
\item Another interesting exploration is how to define a metric score that is better than the current weighted F1 score for Baysian Optimization, to better balance sensitivity and precision.
%This direction is of a more theoretical flavor.
\item There're many problems in assembly areas (including variant calling) that heavily relay on hyper-parameter turing for better performance. Introduce similar frameworks to these problems shall be helpful.
\end{itemize}

% figs too big

%\begin{figure}[!tpb]
%\centerline{\includegraphics[width=0.5\textwidth]{figs/F1.png}}
%\caption{Results of F1.}\label{fig:F1}
%\end{figure}

%\begin{figure}[!tpb]
%\centerline{\includegraphics[width=0.5\textwidth]{figs/sensitivity.png}}
%\caption{Results of Sensitivity.}\label{fig:sens}
%\end{figure}

%\begin{figure}[!tpb]
%\centerline{\includegraphics[width=0.5\textwidth]{figs/precision.png}}
%\caption{Results of Precision.}\label{fig:prec}
%\end{figure}

%\input{split/ack.tex}

\section*{Funding}

This project is funded by NIH R01 Award 1R01HG008164 by NHGRI and NSF CCF Award 1703403.

%%%%%%%%%%%%%%%%%%%%%%%%%%%%%%%%%%%%%%%%%%%%%%%%%%%%%%%%%%%%%%%%%%%%%%%%%%%%%%%%%%%%%
%
%     please remove the " % " symbol from \centerline{\includegraphics{fig01.eps}}
%     as it may ignore the figures.
%
%%%%%%%%%%%%%%%%%%%%%%%%%%%%%%%%%%%%%%%%%%%%%%%%%%%%%%%%%%%%%%%%%%%%%%%%%%%%%%%%%%%%%%

%\bibliographystyle{sty/natbib}
%\bibliographystyle{achemnat}
%\bibliographystyle{plainnat}
%\bibliographystyle{abbrv}
%\bibliographystyle{bioinformatics}
%
%\bibliographystyle{sty/plain}
%\bibliography{split/Document}

\end{document}

\maketitle

%\tableofcontents

%\input{split_supp/supp_dummy.tex}

\section{Subsampling of Alignment}\label{supp_sec:subsample}

\begin{figure}[H] %[h!]
	\centering
	\caption{{Subsampling method across chromosomes of an Alignment}. Right fig is a conceptual illustration of the idea. Chrom A is divided into sub-regions a,b,c,d, chrom B into e,f,g and chrom C into h,i,j,k,l. Among them, b,f,i have high coverage, a,c,e,h,j,l have medium coverage and d,g,k have low coverage. We randomly pick two sub-regions (b,f) from high coverage sub-regions, similarly pick two (e,j) from medium and pick two (d,k) from low coverage sub-regions. Finally we merge randomly picked sub-regions to be sub-sampled alignment.}
	\includegraphics[width=0.99\textwidth]{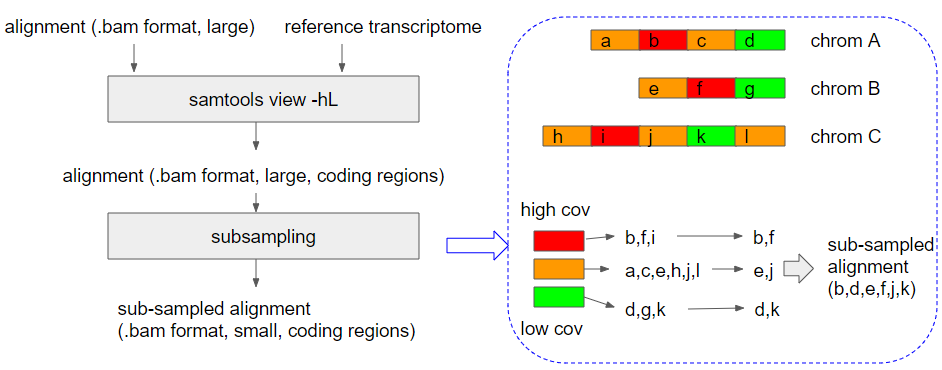}
	\label{fig:subsample}
\end{figure}

To subsample from a large alignment, a simple random sampling is not suitable since the read coverage will be hurt. Instead, particular regions need to be selected. A straight forward way is to retrieve only one chromosome, among multichromosomes, as used in the main paper. Here we describe another way to subsample, as illustrated in Figure \ref{fig:subsample}. Experiments show a similar performance trend as the straight forward method.

Specifically, we first apply reference transcriptome to filter read alignments for only coding regions. Based on read alignments of coding regions, we further divide each chromosome into sub-regions of same span (e.g. $100K$-bp). We calculate the coverage of each sub-region by $coverage = \frac{num\_reads \times read\_len}{span}$. We categorize sub-regions by their coverage (e.g. $5\times$, $10\times$, till $100\times$). For each coverage category, we random pick N (e.g. $N=min(15, {coverage\_category\_size})$) sub-regions. We merge these randomly picked sub-regions from different chromosomes and with various coverages, to be the final sub-sampled alignment.

Take LC dataset as an example. The original large alignment would be reduced by about $97\%$ (e.g. $20G$ to $600M$). The subsampled dataset is of similar size as the one of chromosome 15. BOAssembler tuned this dataset with weighted F1 score, and get sensitivity as $8.4\%$ and precision as $47.9\%$, compared to default's sensitivity as $10\%$ and precision as $39.7\%$.

\section{BO Method}\label{supp_sec:bo}

\subsection{Gaussian Process}
BO typically models the objective function with GP \cite{frazier:2018}. With $k$ acquired data points $\theta_1, ...,\theta_k \in R^d $, and evaluated functions $f(\theta_1), ..., f(\theta_k)$. GP models the prior distribution as multi-variate Normal with mean function $\mu_0(\theta)$ and covariance function $\Sigma_0(\theta, \theta')$. The prior computed given $f(\theta_1), ..., f(\theta_k)$ is: $$f(\theta) \sim N(\mu_0(\theta, \Sigma_0(\theta, \theta')))$$. 

The $\mu_k(\theta)$ and $\Sigma_k(\theta, \theta)$, which refer to the updated mean and covariance functions with datapoints till time $k$, can be estimated explicitly. Given the updated $\mu_k$ and $\Sigma_k$ functions, we can estimate the posterior probablity distribution parameter $\theta_{k+1}$ given datapoints $\theta_{1:k}$ and $f(\theta_{1:k})$. The posterior probablity distribution can be used to decide what parameter to sample in the next iteration.

In this work, we choose mean function to be constant, while the covariance function is estimated via Matern kernel \cite{Snoek:2012}. Kernels compute the distance between points, such that the correlation between a pair of data points are modeled. Matern kernel is defined as:

$$\Sigma_0(\theta, \theta') = \alpha_0 {\frac {2^{1-\nu }}{\Gamma (\nu )}}{\Bigg (}{\sqrt {2\nu }}||\theta - \theta'||{\Bigg )}^{\nu }K_{\nu }{\Bigg (}{\sqrt {2\nu }}||\theta - \theta'||{\Bigg )}$$
Where $K_{\nu }$ is the modified Bessel function. The parameter $\nu$ controls the smoothness of kernel, while $\nu \to \infty$, Matern Kernel converges to RBF kernel; when $\nu=1/2$, Matern Kernel reduces to absolute exponential kernel. $\alpha_0$ controls the variance of the kernel.  

\subsection{Acquisition Function}

Expected Improvement is a widely used acquisition function in BO. When sample a new datapoint $\theta$, and current best parameter $\theta^*$, the improvement is defined as $[f(\theta) - f(\theta^*)]^{+}$. 

The improvement is positive only when $f(\theta)$ is larger than $f(\theta^*)$. Then expected improvement is taken under posterior distributions of $f$ given $\theta_{1:k}$: $$EI_k(\theta) = \E([f(\theta) - f(\theta^*)]^{+}|\theta_{1:k}, f(\theta_{1:k}))$$ 

As expected improvement can be computed in closd-form, we can select the point with largest expected improvement to sample: $\theta_{k+1} = \argmax EI_{k}(\theta)$.

\subsection{Improve performance against BO's randomness}
\begin{figure}[H] %[h!]
	\centering
	\caption{Learning Curve of Different BO runs. 40 itertions with 10 runs shown}
	\includegraphics[width=0.99\textwidth]{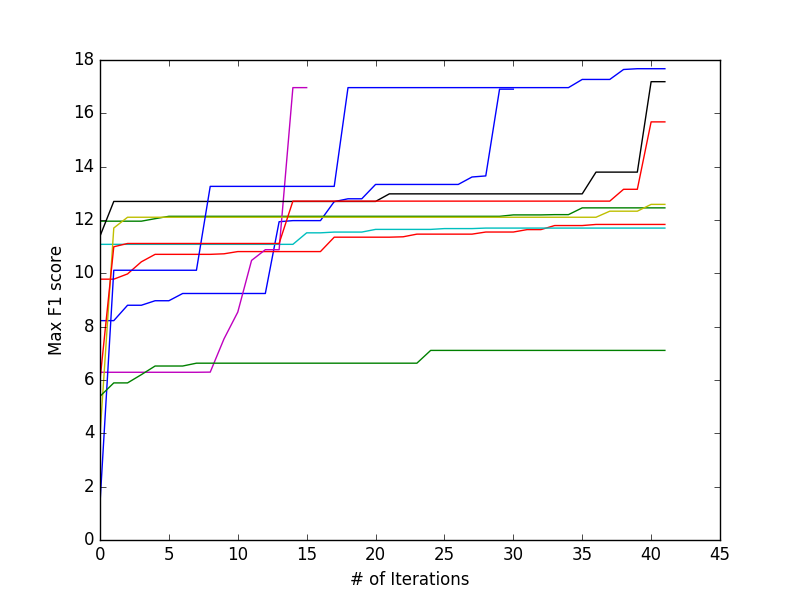} 
	\label{fig:sup_learning_curve}
\end{figure}

BO's randomness come from the acquisition selection with posterior, which lead to undesired local optimum parameters lead to staled performance. To avoid BO getting stuck, one practical suggestion is to conduct multiple BO runs, shown in Figure \ref{fig:sup_learning_curve}. 

Learning curve shows the best performance parameters acquired so far by BO. Some learning curves converge to better performance, while other learning curves saturate at some local optimum. With single run, BO might get stuck, but with multiple runs, the performance will improve. When applying BO in practice, we suggests to run multiple BO to ensure better performance.

\subsection{Avoid BO Overfitting by cross-validation}
BO tunes parameters $\theta$ on small dataset, and test the performance on large dataset, which draws concern on overfitting. BO overfitting means parameter $\theta$ is over-optimized for the small dataset, while the performance on large dataset degrades.

We can use additional validation dataset to avoid BO overfitting. We take two small dataset of reads, $D_{train}$ is for training and $D_{val}$ for validation. $f(.)$ is optimized solely on $D_{train}$ to get $\theta_t$, while for each iteration the $f(D_{val}, \theta_t)$ is evaluated. After all iterations, the best parameter evaluated on $D_{val}$ is returned. Use the cross-validation on BO, we can potentially avoid overfitting parameter to training dataset.%\textbf{Some experiment to support?}

\section{Hyper-Parameter}

Here we offer a screen shot of the hyper-parameter configuration for Stringtie.

\begin{figure}[H] %[h!]
	\centering
	\caption{{Hyper-parameter Configuration of Stringtie in BOAssembler}}
	\includegraphics[width=0.9\textwidth]{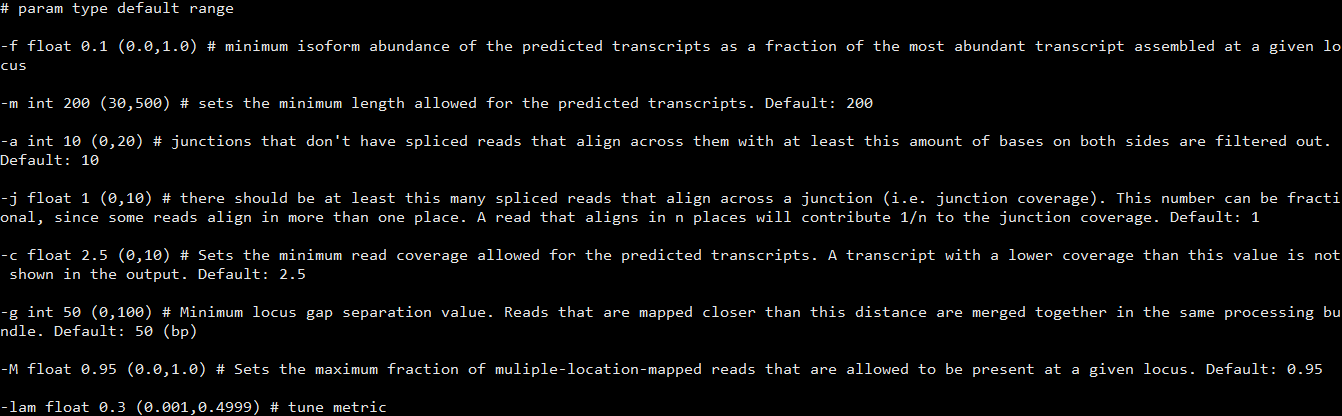}
	\label{supp_fig:hyperparam}
\end{figure}